%% file: main.tex
\documentclass[runningheads]{llncs}
\usepackage[T1]{fontenc}
\usepackage{graphicx,verbatim}
\usepackage{microtype}
\usepackage{amsmath}
\usepackage{amssymb}
\usepackage{bm}
\usepackage{algorithm}
\usepackage{algpseudocode}
\usepackage{placeins}

\begin{document}
%
% \title{Automated subtype discovery in connectome-based disease progression modeling of Parkinson's disease}

\title{Discovering Subtypes of Neurodegenerative Progression with a Scalable Connectome-Constrained Dynamic Model}

\titlerunning{Subtype Discovery in Connectome-based DPM}
% If the paper title is too long for the running head, you can set
% an abbreviated paper title here
%

\authorrunning{D. Semchin et al.}

\begin{comment}  %% Removed for anonymized MICCAI submission
\author{First Author\inst{1}\orcidID{0000-1111-2222-3333} \and
Second Author\inst{2,3}\orcidID{1111-2222-3333-4444} \and
Third Author\inst{3}\orcidID{2222--3333-4444-5555}}
%

% First names are abbreviated in the running head.
% If there are more than two authors, 'et al.' is used.
%
\institute{Princeton University, Princeton NJ 08544, USA \and
Springer Heidelberg, Tiergartenstr. 17, 69121 Heidelberg, Germany
\email{lncs@springer.com}\\
\url{http://www.springer.com/gp/computer-science/lncs} \and
ABC Institute, Rupert-Karls-University Heidelberg, Heidelberg, Germany\\
\email{\{abc,lncs\}@uni-heidelberg.de}}

\end{comment}

\author{Daniel Semchin\inst{1} \and
Emile d'Angremont\inst{2} \and
Hao Ding\inst{1} \and
Alan Antar\inst{1} \and
Marco Lorenzi\inst{3} \and
Konstantinos Arfanakis\inst{1} \and
Ysbrand van der Werf\inst{2} \and
Paul Thompson\inst{4} \and
Boris Gutman\inst{1}}

\institute{
Illinois Institute of Technology, Chicago, IL, USA \and
Amsterdam UMC, Amsterdam, The Netherlands \and
Inria, Universit\'e C\^ote d'Azur, Sophia Antipolis, France \and
University of Southern California, Los Angeles, CA, USA
}
  
\maketitle              % typeset the header of the contribution
\begin{abstract}
%Neurodegenerative diseases exhibit substantial heterogeneity in symptomatology, with patients showing varying degrees of cognitive decline and motor dysfunction. 
Parkinson's disease is clinically and biologically heterogeneous, yet its spatiotemporal progression remains poorly characterized. We present a connectome-constrained disease progression model that jointly estimates subject-specific disease time and data-driven subtypes from longitudinal morphometry. Applied to 85 imaging and clinical biomarkers from the Parkinson's Progressive Markers Initiative (PPMI) cohort, the model recovers four morphologically distinct progression subtypes. We validate the model on a hold-out cross-sectional dataset and benchmark it against SuStaIn under a matched training and validation protocol. Only our method recovers subtypes that correspond significantly to clinical motor subtypes and genetic variants of Parkinson's Disease. 

% Both models recover a consistent subcortical-versus-cortical organizing structure, and neither shows strong separation by motor phenotype.

% Authors must provide keywords and are not allowed to remove this Keyword section.

\end{abstract}
\input{chapters/1-introduction.tex}

\input{chapters/2-Method.tex}

\input{chapters/3-data-experiments.tex}

\input{chapters/4-results.tex}

\input{chapters/5-future-plans-conclusion.tex}

\input{chapters/6-acknowledgments.tex}

% \printbibliography
\bibliographystyle{splncs04}
\bibliography{MIC2026_bib}

\end{document}

%% file: chapters/1-introduction.tex
\section{Introduction}

Models of disease progression (DPM) offer a means to simulate treatment effect – sometimes called “digital twins” – and to estimate pre-symptomatic time of disease onset. Conventional DPM’s in the context of neuroimaging infer long-term biomarker trajectories by “stitching together” longitudinal observations collected over a relatively short period from large sample sets. The problem requires simultaneous estimation of the canonical trajectory as well as each patient’s disease score, typically treated as the free “time” variable mapping the patient to the trajectory. Established models include discrete-time \cite{1-EBM,2-SusTain} and continuous-time variants \cite{5_BrLP,4_DP_Most}. Several continuous-time DPMS use non-linear dynamics constrained by structural connectivity \cite{6_NDM,7_ACP,8_NODE_progression,15_TauFlowNet,16_COMIND}, providing a mechanistic framework that mirrors "prion-like" transsynaptic transmission of disease agents. 
A related challenge in progression modeling is disease heterogeneity. Patients with in the same general diagnostic category often follow distinct progression patterns, motivating subtype discovery within DPMs. Subtyping progression models exist \cite{2-SusTain,4_DP_Most}, but are generally either not scalable to large numbers of biomarkers or lack the generative mechanistic power of network-based models.
As highlighted in recent reviews \cite{13_Young_DPM_review,14_Moravveji_DPM_review}, a critical gap remains in our current modeling arsenal: the lack of approaches that simultaneously leverage the brain's connectivity structure, scale to large numbers of brain regions, accommodate disease heterogeneity, and provide mechanistic interpretability over extended time ranges. In this work, we extend the recently proposed network-constrained COMIND model \cite{16_COMIND}  to enable simultaneous modeling of nonlinear network dynamics and disease subtype identification. Our proposed method models neurodegeneration as a system of coupled logistic-diffusion trajectories constrained by structural connectivity, and performs automatic subtype discovery by inferring distinct spatial patterns of regional pathology sources from longitudinal imaging data.

%% file: chapters/2-Method.tex
\section{Methods}

A number of the proposed progression models have used logistic evolution as the basic model-building element. The idea is appealing as it mirrors our intuition that neurodegeneration has a gradual onset, a period of rapid advance, and a saturation point. Indeed, the sigmoid is a solution to the logistic ODE $\dot{x}=k(1-x)x$, a dynamic in which the rate of accumulation is a product of self-supply and capacity. Our motivation is to incorporate network dynamics into this idea, while maintaining the asymptotic properties of the univariate logistic ODE, and to identify disease subtypes characterized by distinct patterns of regional pathology.

\subsection{Dynamic Propagation Model}

We begin by considering a dynamic $p$-dimensional system of accumulating neurodegenerative effects with a static transition matrix $K\in\mathbb{R}^{p\times p}$ defined in some way by brain connectivity. As in a standard logistic differential equation, we define the rate of neurodegenerative accumulation as a product of the present state of neurodegeneration and remaining capacity. This can be expressed as 

\begin{equation}
\label{naive_logistic}
    \frac{d\mathbf{x}}{dt} = [I-D(\mathbf{x})]K\mathbf{x},\;\; \mathbf{x}\in[0,1]^p
\end{equation}

Here, $D(\mathbf{x})$ is the diagonal matrix with entries in $\mathbf{x}$. We consider external sources of neurodegeneration as constant, again contributing to the rate of regional accumulation in proportion to regional capacity. 

\begin{equation}
\label{logistic_forcing}
    \frac{d\mathbf{x}}{dt} = [I-D(\mathbf{x})][K\mathbf{x} + \mathbf{f}],\;\; \mathbf{x}\in[0,1]^p, \;\; \mathbf{f}\in[0,\infty)^p, 
\end{equation}

where $\mathbf{f}$ represents the constant "forcing" term in the ODE. For the $k^\text{th}$ individual regional biomarker, this is equivalent to
\begin{equation*}
    \dot{x}^k(t)=(1-x^k)(\sum_mK_{km}x^m + f^k),
\end{equation*}
The transition matrix combines two parameters governing regional connectivity: network mediated propagation along the connectome, and local, connectivity-independent accumulation. Formally,
\begin{equation}
\label{eq:diagonal_transition}
    K = s_tK^* + \mathrm{diag}(\boldsymbol{\kappa}),
\end{equation}
where $s_t$ is a scalar timescale governing the strength of network-mediated propagation and $\boldsymbol{\kappa}\in\mathbb{R}^p$ is a vector of region-specific self-propagation rates. Finally, to construct canonical trajectories of neurodegeneration we must scale solutions of the dynamic system to the natural scale of imaging biomarkers. This adds a scaling vector $\mathbf{s}\in\mathbb{R}^p$ to our trajectory parameter set. Suppose $\mathbf{x}(t)$ is the solution to \eqref{logistic_forcing}. The predicted biomarker value at time $t$ is then $\mathbf{\hat{y}}(t)= \mathbf{s}\odot \mathbf{x}(t)$, where $\odot$ is element-wise multiplication. 

To account for population heterogeneity, we define $Z$ subtypes. Each subtype $z \in \{1,\ldots,Z\}$ has its own forcing term $\mathbf{f}_z$, representing a distinct spatial pattern of external pathology sources, 
while $s_t$ and $\mathbf{s}$ are shared globally across subtypes. Each subject receives a hard assignment to the subtype whose predicted trajectory yields the lowest reconstruction error at their estimated 
disease time. The full parameter set is $\theta = \{s_t, \kappa, \mathbf{s}, \mathbf{f}_1, \ldots, \mathbf{f}_Z\}$, with $2p+1$ global parameters and $Z \cdot p$ subtype-specific forcing parameters.

\subsection{Modeling Subject-Specific Time-Shift}
As is common practice, we assume a Gaussian i.i.d. noise model. For the $k^{\text{th}}$ biomarker $y_{ij}^k$ measured in subject $i$ at timepoint $j$, biomarker likelihood is expressed as 

\begin{equation}
\label{single_biomarker_like}
    P(y_{ij}^k \mid \theta, \beta_i, z_i) = \mathcal{N}(y_{ij}^k \mid 
    \hat{y}^k(t_{ij}+\beta_i;\theta_{z_i}),\sigma_k).
\end{equation}

Here, $\beta_i$ is the subject-specific "disease-time" or time-shift at the initial scan, and $t_{ij}$ is the time of scan at visit $j$, given that $t_{i0}=0$. To discourage subtypes from separating primarily along $\beta_i$ rather than underlying progression pattern, we add a Jensen-Shannon divergence penalty weighted by $\lambda_{\text{jsd}}$ that encourages distributions of time-shifts to be similar across subtypes. 

\subsection{Biomarker Trajectory Conditioning and Likelihood Estimation}

To represent the fact that brain function is largely self-contained and locally protected from external inputs, we assume that the number of brain regions affected by $\mathbf{f}$ is small and therefore $\mathbf{f}$ is sparse. We place the same sparsity assumption on the self-connection term $\boldsymbol{\kappa}$, reflecting the expectation that local connectivity-independent accumulation is relevant in only a subset of regions. To prevent the timescale from collapsing to zero or diverging, we place a log-normal prior on $s_t$ around a prior value $s_{\text{pr}}$. The complete prior on model parameter is 

\begin{equation}
\label{theta_prior}
    P(\theta) \propto \exp\!\left[-\lambda_f\sum_{z=1}^{Z}
    \|\mathbf{f}_z\|_1 - \lambda_\kappa\|\boldsymbol{\kappa}\|_1
    - \frac{\lambda_s}{2}(\log \frac{s_t}{s_{\text{pr}}})^2 
    - \lambda_{\mathbf{s}}\|\mathbf{s}\|_2^2\right]
\end{equation}

Combining the terms in \eqref{single_biomarker_like} and \eqref{theta_prior}, we have the posterior for $\theta$ and $\beta$:
\begin{equation}
\label{posterior}
    P(\theta,\beta,\mathbf{z} \mid \mathbf{y}) = 
    \prod_{ijk} P(y_{ij}^k \mid \theta,\beta_i,z_i) \times P(\theta)
\end{equation}

Subtype priors $P(z_i) = 1/Z$ are assumed uniform.

\subsection{Model Estimation}
We estimate model parameters and subject time-shifts using a generalized Expectation-Maximization approach. %Beyond priors on model parameters, we penalize the Jensen-Shannon divergence (JSD) of $\beta$ distributions across subtypes; this prevents subtyping from becoming a surrogate for clustering disease stage. 
To reduce sensitivity to local optima, the algorithm is run from multiple random initializations of $\beta$ and subtype assignments $\mathbf{z}$, retaining the solution with the lowest training loss. The number of subtypes $Z$ is selected using the Bayesian Information Criterion (BIC), 
% \begin{equation}
% \text{BIC} = -2\ln\hat{L} + \nu \ln n,
% \end{equation}
penalizing model complexity to avoid selecting an unnecessarily large number of subtypes. The effective parameter count $\nu$ includes the $p$ entries of $\mathbf{s}$, the global timescale $s_t$, the number of nonzero entries of $\boldsymbol{\kappa}$, and the number of nonzero forcing term entries across all subtypes. The observation count $n$ is the total number of scalar measurements, i.e. the sum over all subjects of visits times biomarkers. The log likelihood term is the sum of per-biomarker squared residuals normalized by the observed variance of each biomarker. 

\begin{algorithm}[!h]
\caption{SubtypingEM}
\label{alg:subtyping_em}
\begin{algorithmic}[1]
\Require observations $\mathbf{y}$, 
         connectome $K^*$, number of subtypes $Z$
\Ensure $\theta = \{s_t, \boldsymbol{\kappa}, \mathbf{s}, \mathbf{f}_1,\ldots,\mathbf{f}_Z\}$, 
        $\beta$, $\mathbf{z}$
\State Initialize $\beta$, $\mathbf{z}$ randomly; initialize 
       $\mathbf{f}_z$ for each subtype
\While{not converged}
    \State Update $\mathbf{s}$, $s_t$, $\boldsymbol{\kappa}$ with $\beta$, 
           $\mathbf{z}$, $\{\mathbf{f}_z\}$ fixed \hfill (L-BFGS)
    \State $z_i \leftarrow \arg\min_z \; 
           \mathcal{L}_{\text{recon}}(\beta_i, \mathbf{y}_i; \theta_z)$ 
           for each subject $i$
    \State Update each $\mathbf{f}_z$ using subjects 
           with $z_i = z$ \hfill (L-BFGS)
    \State  Update all $\beta_i$ jointly 
           via \eqref{posterior} with JSD regularization \hfill (L-BFGS)
\EndWhile
\end{algorithmic}
\end{algorithm}

\FloatBarrier

%% file: chapters/3-data-experiments.tex
\section{Data and Experiments}

\subsubsection{Synthetic Data generation}
\begin{figure}[t]
    \centering
    \includegraphics[width=\textwidth]{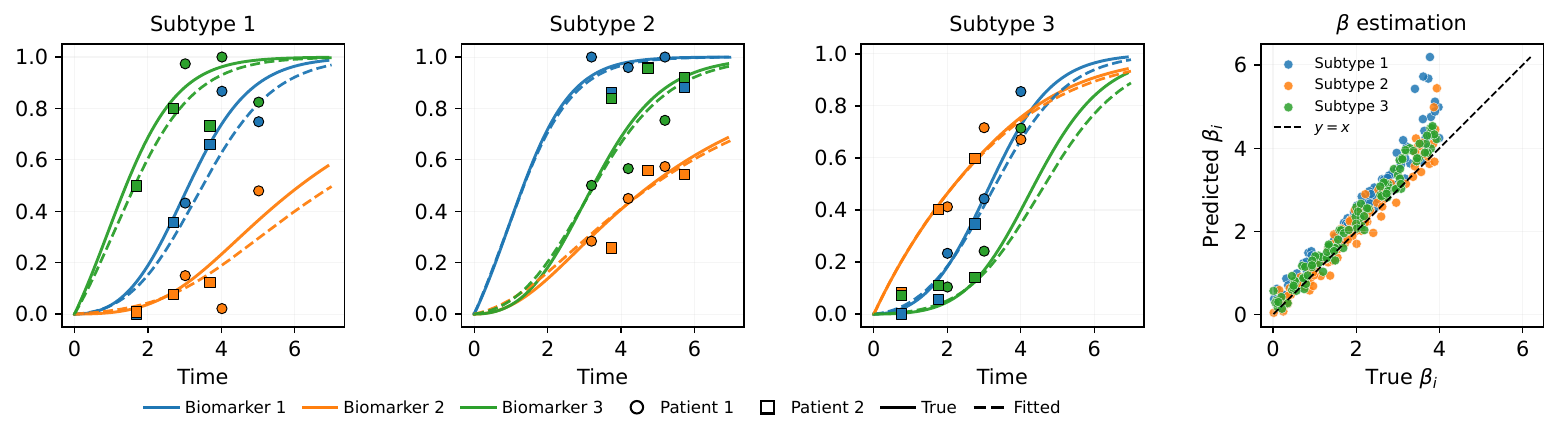}
    \caption{Synthetic validation and trajectory recovery for three ground-truth subtypes. \textbf{Left three: }Trajectory recovery. Each column corresponds to one subtype. Noisy patient observations with fitted trajectories (dashed) against ground-truth (solid). \textbf{Right: }Synthetic validation.  predicted vs.\ true subject time-shifts $\beta$. Points are centered on the $\hat{\beta}=\beta$ line, indicating unbiased recovery. The assigned subtype co-clustered with true subtype nearly perfectly (Adjusted Rand Index $= 0.96$). }
    \label{fig:synthetic_scatter_validation}
\end{figure}

We validated COMIND on synthetic data generated from three ground-truth subtypes, each governed by \eqref{logistic_forcing} with $p=3$ biomarkers and a shared, fixed connectivity matrix $K$ and shared scalar coupling $\alpha=0.2$. Subtypes differ only in their forcing terms $\mathbf{f}$, with each subtype assigned a single nonzero forcing entry at a distinct biomarker index; this construction directly tests whether EM can recover subtype-specific forcing patterns from a shared network structure. For each subtype, 100 subjects were simulated with 3 observations each, sampled at yearly intervals along a 7-year disease-time axis, with Gaussian observation noise ($\sigma=0.1$) added to all biomarkers. We additionally generated a privileged clinical feature per observation as $a(\beta_i + t_{ij} + \mathcal{N}(0,1)) + b$, with slope $a\in[1,5]$ and bias $b\in[0,10]$, to test the model's clinical-informed initialization pathway: subject time-shifts $\beta$ are initialized via a linear mixed-effects model fit to these clinical scores, and forcing terms are initialized from the leading eigenvectors of $K^*$. Models were fit for $Z\in\{2,3,4\}$ with 3-fold cross-validated hyperparameter selection, and $Z$ was chosen by BIC. Trajectory and sample generation for each subtype are shown in Fig.~\ref{fig:synthetic_scatter_validation}.

\subsection{Parkinson's Disease Imaging and Clinical Data}

The PPMI study employs a standardized high-resolution 3D T1-weighted volumetric sequence (MP-RAGE/IR-FSPGR) acquired on 3T scanners across ~50 international sites (sagittal $1.0\times1.0\times1.0$ mm isotropic scan, $256^3$ FOV). We used 313 PPMI subjects with Parkinson's Disease, yielding 622 total observations. T1-weighted scans were processed using FreeSurfer 7.0 to harmonize cortical and subcortical parcellations (Desikan-Killiany atlas, DK) across the time points, and residual site effects were removed using ComBat harmonization. We use cortical thickness measures 68 DK regions and 14 subcortical regions, expressed as deviation Z-scores relative to a normative reference cohort as our primary biomarker. The Z-scores were floored at $\pm 6$ SD, sign-flipped so pathology corresponds to increasing biomarker value, and each biomarker was translated by its own minimum so all values are non-negative, as required by the model. Additional clinical measures included Montreal Cognitive Assessment (MoCA), Tremor-dominant sub-score (TD-Score) and Postural Instability-Gait Disorder Subscore (PIGD-score) from MDS-UPDRS-III symptom-specific items. Lower MoCA and higher TD/PIGD scores imply greater symptom severity. 

\subsection{Connectome Model}

We used the precomputed structural connectivity matrix from the IIT Human Brain Atlas v5.0 \cite{iit_atlas}, which was derived from whole-brain tractography on a HARDI template. Tractography was performed using Anatomically-Constrained Tractography (ACT) and filtered with the Spherical-deconvolution Informed Filtering of Tractograms (SIFT) algorithm. Streamlines were mapped onto the 82 cortical and subcortical regions of the Desikan-Killiany atlas to produce a group-level connectivity matrix. We retained only the top 10\% of connections by streamline count, verified that no regions became disconnected after thresholding, and normalized the resulting matrix by its median row sum. Self-connections were set to zero. Zero padding was added to the connectome to accomodate clinical measures as disconnected regions in the model.

\subsection{Parkinson's Disease Experiments}

We split subjects by number of visits: 126 patients with longitudinal follow-up (435 observations) formed the training set, and 187 with a single cross-sectional scan formed the hold-out validation set. We performed grid search over hyperparameters $\lambda_f$, $\lambda_{\text{scalar}}$, $\lambda_{\kappa}$, and $\lambda_{\text{jsd}}$, selecting the combination yielding the smallest BIC.

%% file: chapters/4-results.tex
\section{Results}

\subsection{Synthetic results}

BIC correctly identified $Z=3$ as the optimal number of subtypes over candidates $Z \in \{2,3,4\}$. Subtype-specific forcing terms were recovered accurately: in each case the dominant nonzero entry was correctly identified at the true biomarker index, with fitted values of $0.317$, $0.302$, and $0.249$ against ground truth values of $0.3$. The self-connection term $\boldsymbol{\kappa}$ was likewise recovered well: the true zero entry was correctly estimated as exactly zero by the sparsity prior, and the two nonzero entries (true $0.651$, $0.604$) were recovered as $0.726$ and $0.650$. Subtype classification achieved an Adjusted Rand Index of $0.96$, and mean absolute time-shift error was $0.345 \pm 0.317$ years on a 7-year disease axis across $300$ subjects. Subtype classification and time-shift recovery are shown in Fig.~\ref{fig:synthetic_scatter_validation}.

\subsection{Parkinson's Disease Model}
\begin{figure}[t]
    \centering
    \includegraphics[width=\linewidth]{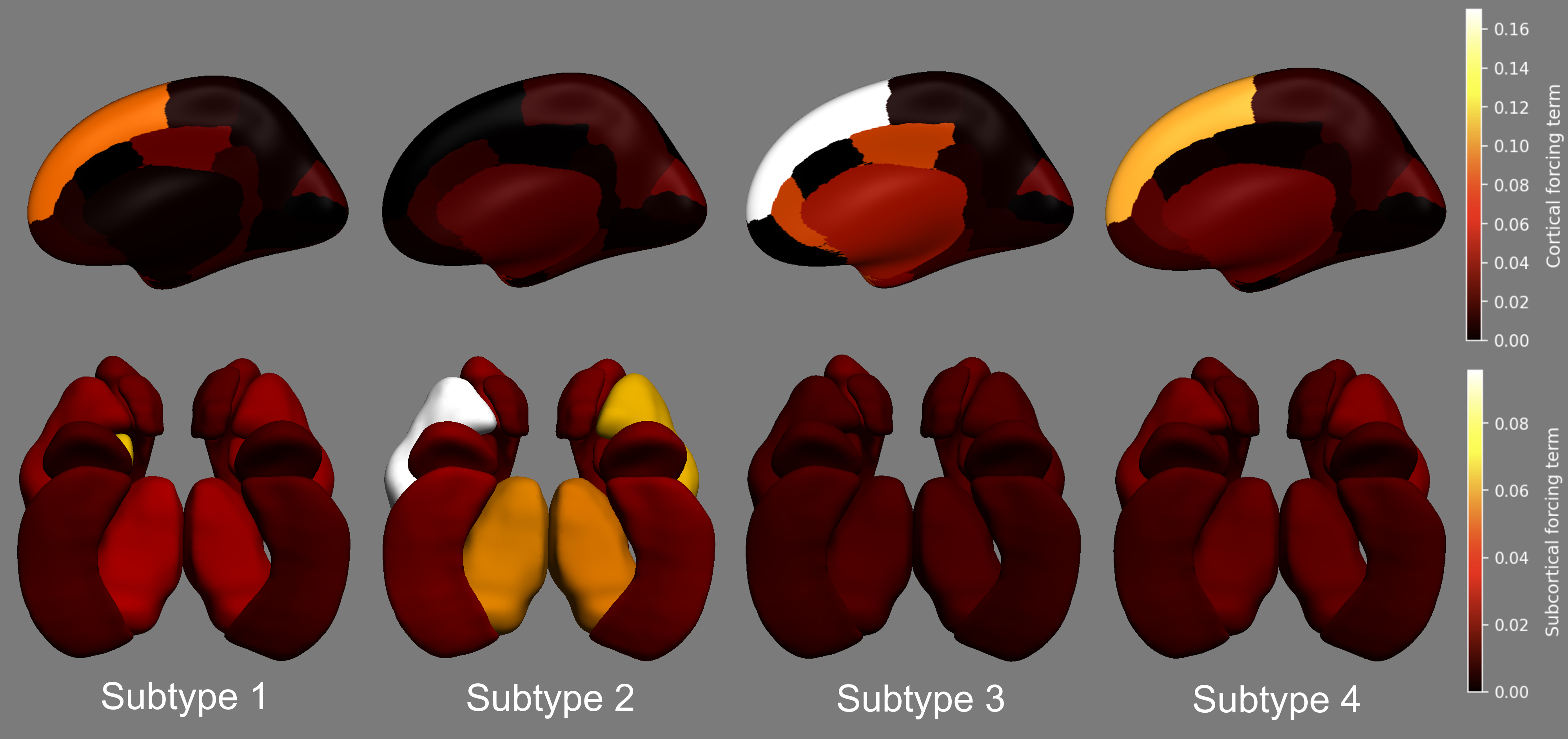}
    \caption{Right-medial and axial view of brain region forcing terms}
    \label{fig:progression_maps}
\end{figure}

\begin{figure}
    \centering
    \includegraphics[width=\linewidth]{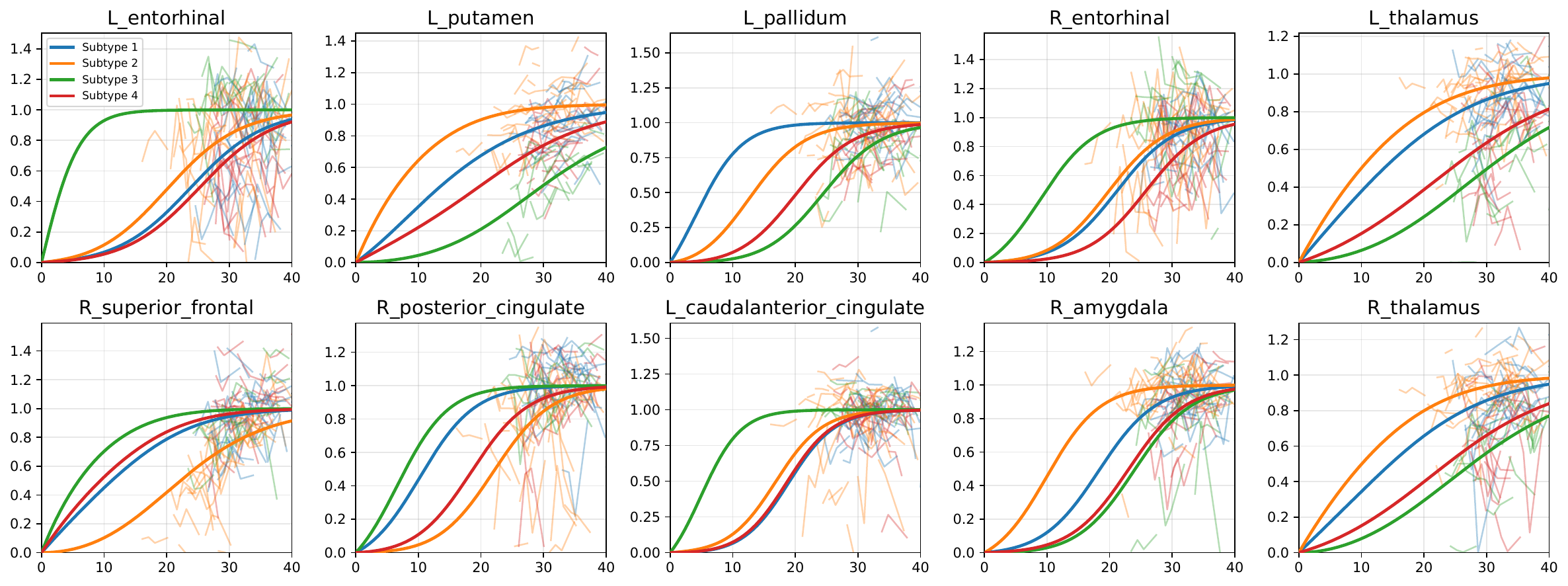}
    \caption{Subtype-specific cortical thickness trajectories for the ten regions showing the greatest pairwise $\ell_2$ difference between subtype curves. Observed subject trajectories are overlaid on each subtype's fitted curve.
    }
    \label{fig:spaghetti}
\end{figure}

BIC selected $Z=4$ subtypes over candidates $Z \in \{2,3,4\}$. Subtype-specific trajectories for the regions showing the greatest pairwise $\ell_2$ difference between subtype curves are shown in Fig.~\ref{fig:spaghetti}. Across the subtypes, different subsets of biomarkers drive progression, denoted by the magnitude of forcing terms across them. Subtype 1 is primarily driven by TD score, alongside pallidal and thalamic forcing terms. Subtype 2 is jointly driven by PIGD score and MCATOT, the two largest forcing terms in the model, together with strong bilateral putamen and thalamus forcing (Fig.~\ref{fig:progression_maps}). Subtype 3 is more cortically driven, primarily by entorhinal cortex, superior frontal cortex, and anterior cingulate. Subtype 4 is jointly driven by high TD and PIGD forcing, with greater forcing term in the superior frontal and temporal pole regions. 

% Fig.~\ref{fig:clinical} shows the relationship between discovered subtypes and clinical phenotype. The contingency table comparing subtype assignment to TD/PIGD classification reveals that subtype 2 is predominantly PIGD-dominant with no indeterminate cases, while subtypes 0 and 1 show a more mixed phenotypic distribution. Notably, subtype 2 also exhibits the largest PIGD regression coefficient and the only positive MoCA coefficient across subtypes, suggesting a clinically coherent profile. Notably, trajectories showing greatest differences in their staging and rate of gray matter loss among the discovered subtypes are in regions primarily associated with early stages of cognitive decline. This corresponds to the known clinical association between the PIGD type and cognitive decline, also seen in our results. This separation emerged without the use of clinical priors, supporting the interpretation that the model captures morphologically distinct progression patterns that correspond to known clinical phenotypes. % Given the small size of subtype 2 ($n=21$), findings specific to this subtype should be approached with caution.

% \begin{figure}
%     \centering
%     \includegraphics[width=\linewidth]{chapters/images/comind/comind_clinical_trajectories.pdf}
%     \caption{Subtype-specific clinical measurement trajectories for clinical biomarkers: MoCA (MCATOT), TD Score and PIGD score.
%     }
%     \label{fig:clin_curve}
% \end{figure}

\begin{figure}
    \centering
    \includegraphics[width=\linewidth]{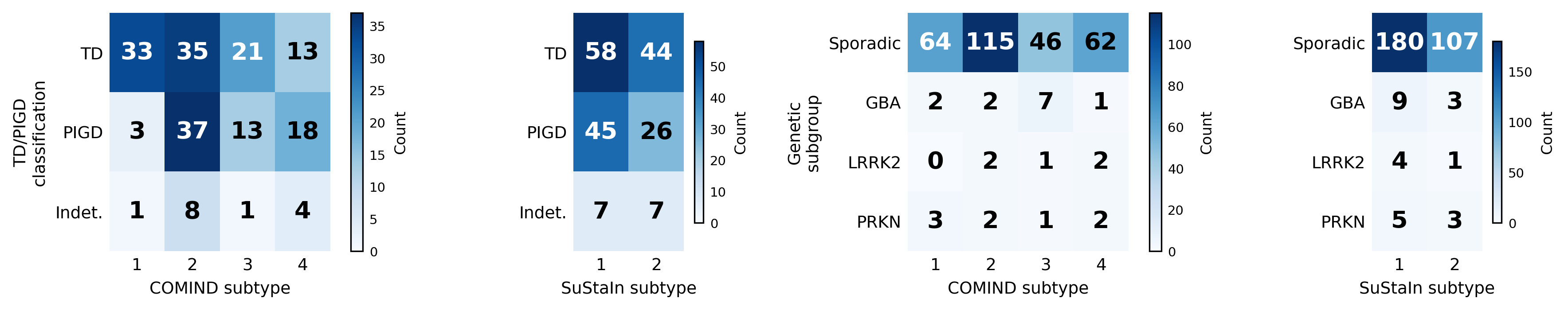}
    \caption{Contingency tables of discovered subtype assignment versus clinical and genetic groupings. From left to right: COMIND and SuStaIn TD/PIGD/Indeterminate motor phenotype classification, held-out validation cohort ($n=187$); COMIND and SuStaIn versus genetic subgroup (GBA, LRRK2, PRKN, sporadic PD), full cohort ($n=312$).}
    \label{fig:sustain_comind_summary}
\end{figure}

\subsection{Comparison to SuStaIn}
To contextualize COMIND against a purely data-driven staging model, we fit SuStaIn~\cite{2-SusTain} (pySuStaIn~\cite{11_pySuStaIn}) using the same underlying z-scored data, aggregated to 13 regions and 3 clinical scores for computational tractability, and the same training/validation split as COMIND; cross-validation favored a two-subtype solution, selected via pySuStaIn's built-in cross-validation information criterion (CVIC), resembling the \emph{Cortical} and \emph{Subcortical} staging patterns reported by Shawa et al.~\cite{9_Shawa} (Fig.~\ref{fig:sustain_comind_summary}). We compared subtype assignments from both methods against two independent groupings: TD/PIGD motor phenotype~\cite{10_stebbins2013identify} and genetic subgroup (GBA, LRRK2, PRKN, sporadic PD), using $\chi^2$ tests of independence (Fig.~\ref{fig:sustain_comind_summary}). On the hold-out validation cohort ($n=187$), SuStaIn's two subtypes showed no significant association with TD/PIGD classification ($\chi^2=1.22$, $df=2$, $p=0.543$), whereas COMIND's four subtypes showed a significant association ($\chi^2=27.33$, $df=6$, $p<0.001$). The same pattern held for genetic subgroup across the full cohort ($n=312$): SuStaIn showed no significant association ($\chi^2=1.35$, $df=3$, $p=0.717$), while COMIND showed a significant association ($\chi^2=17.80$, $df=9$, $p=0.038$). These results suggest COMIND's connectome-constrained subtypes capture clinically and genetically meaningful structure.

%% file: chapters/5-future-plans-conclusion.tex
\section{Conclusion}
We introduced a connectome-constrained progression model that jointly recovers subject-level disease time and morphologically distinct subtypes from cortical and subcortical trajectories. COMIND identified four progression subtypes in the PPMI cohort. These subtypes correspond significantly to genetic variants of PD prescribed by the GBA, LRRK2 and PRKN variants as well as clinical motor subtypes in a holdout dataset. A matched comparison to SuStaIn showed that both models recover a consistent subcortical-versus-cortical organizing structure despite their differing formulations. Validation on independent cohorts such as ENIGMA-PD \cite{12_ENIGMA_PD} and incorporation of dynamically evolving connectome structure remain directions for future work.

% We have presented an extension of COMIND that performs automatic subtype discovery in connectome-based disease progression modeling. Each subtype is characterized by a distinct spatial pattern of external pathology sources, while global connectivity and scaling parameters are shared. The number of subtypes is selected by BIC, avoiding arbitrary model complexity. On synthetic data the model accurately recovered subtype-specific parameters, subject time-shifts, and subtype assignments. On PPMI, three morphological subtypes were discovered from cortical thickness data without clinical supervision, yet the discovered subtypes show meaningful alignment with established motor phenotypes. Future work includes validation on independent cohorts such as ENIGMA-PD \cite{12_ENIGMA_PD}, incorporation of dynamic connectome structure, and extension to multimodal biomarkers.

%% file: chapters/6-acknowledgments.tex
\section{Acknowledgments}

This work was supported by the Michael J Fox Foundation grant MJFF-021683, Multimodal Dynamic Modeling and Prediction of Parkinsonian Symptom Progression

%% file: MIC2026_bib.bib
@article{1-EBM,
   author = {Fonteijn, H. M. and Modat, M. and Clarkson, M. J. and Barnes, J. and Lehmann, M. and Hobbs, N. Z. and Scahill, R. I. and Tabrizi, S. J. and Ourselin, S. and Fox, N. C. and Alexander, D. C.},
   title = {An event-based model for disease progression and its application in familial {Alzheimer's} disease and {Huntington's disease}},
   journal = {Neuroimage},
   volume = {60},
   number = {3},
   pages = {1880-9},
   ISSN = {1095-9572 (Electronic)
1053-8119 (Linking)},
   DOI = {10.1016/j.neuroimage.2012.01.062},
   url = {https://www.ncbi.nlm.nih.gov/pubmed/22281676},
   year = {2012},
   type = {Journal Article}
}

@article{2-SusTain,
   author = {Young, A. L. and Marinescu, R. V. and Oxtoby, N. P. and Bocchetta, M. and Yong, K. and Firth, N. C. and Cash, D. M. and Thomas, D. L. and Dick, K. M. and Cardoso, J. and van Swieten, J. and Borroni, B. and Galimberti, D. and Masellis, M. and Tartaglia, M. C. and Rowe, J. B. and Graff, C. and Tagliavini, F. and Frisoni, G. B. and Laforce, R., Jr. and Finger, E. and de Mendonca, A. and Sorbi, S. and Warren, J. D. and Crutch, S. and Fox, N. C. and Ourselin, S. and Schott, J. M. and Rohrer, J. D. and Alexander, D. C.},
   title = {Uncovering the heterogeneity and temporal complexity of neurodegenerative diseases with Subtype and Stage Inference},
   journal = {Nat Commun},
   volume = {9},
   number = {1},
   pages = {4273},
   ISSN = {2041-1723 (Electronic)
2041-1723 (Linking)},
   DOI = {10.1038/s41467-018-05892-0},
   url = {https://www.ncbi.nlm.nih.gov/pubmed/30323170},
   year = {2018},
   type = {Journal Article}
}

@inproceedings{4_DP_Most,
   author = {Viani, A. and Gutman, B. A. and d’Angremont, E. and Lorenzi, M.},
   title = {Disease Progression Modelling and Stratification for Detecting Sub-trajectories in the Natural History of Pathologies: Application to Parkinson’s Disease Trajectory Modelling},
   series = {Medical Image Computing and Computer Assisted Intervention – MICCAI 2024 Workshops},
   publisher = {Springer Nature Switzerland},
   pages = {3-14},
   ISBN = {978-3-031-84525-3},
   type = {Conference Proceedings}
}

@InProceedings{5_BrLP,
        author = { Puglisi, L. and Alexander, D. C. and Ravì, D.},
        title = { { Enhancing Spatiotemporal Disease Progression Models via Latent Diffusion and Prior Knowledge } },
        booktitle = {proceedings of Medical Image Computing and Computer Assisted Intervention -- MICCAI 2024},
        year = {2024},
        publisher = {Springer Nature Switzerland},
        volume = {LNCS 15002},
        month = {October},
        page = {173 -- 183}
}

@article{6_NDM,
   author = {Raj, A. and LoCastro, E. and Kuceyeski, A. and Tosun, D. and Relkin, N. and Weiner, M.},
   title = {Network Diffusion Model of Progression Predicts Longitudinal Patterns of Atrophy and Metabolism in {Alzheimer's} Disease},
   journal = {Cell reports},
   volume = {10},
   number = {3},
   pages = {359-369},
   ISSN = {2211-1247},
   DOI = {10.1016/j.celrep.2014.12.034},
   url = {https://www.ncbi.nlm.nih.gov/pubmed/25600871},
   year = {2015},
   type = {Journal Article}
}

@article{7_ACP,
   author = {Garbarino, S and Lorenzi, M},
   title = {Investigating hypotheses of neurodegeneration by learning dynamical systems of protein propagation in the brain},
   journal = {NeuroImage},
   volume = {235},
   pages = {117980},
   ISSN = {1053-8119},
   DOI = {https://doi.org/10.1016/j.neuroimage.2021.117980},
   url = {https://www.sciencedirect.com/science/article/pii/S1053811921002573},
   year = {2021},
   type = {Journal Article}
}

@article{8_NODE_progression,
   author = {Abi Nader, C. and Ayache, N. and Frisoni, G. B. and Robert, P. and Lorenzi, M.},
   title = {Simulating the outcome of amyloid treatments in {Alzheimer's} disease from imaging and clinical data},
   journal = {Brain Communications},
   volume = {3},
   number = {2},
   pages = {fcab091},
   ISSN = {2632-1297},
   DOI = {10.1093/braincomms/fcab091},
   url = {https://doi.org/10.1093/braincomms/fcab091},
   year = {2021},
   type = {Journal Article}
}

@article{9_Shawa,
    author = {Shawa, Zeena and Shand, Cameron and Taylor, Beatrice and Berendse, Henk W. and Vriend, Chris and van Balkom, Tim D. and van den Heuvel, Odile A. and van der Werf, Ysbrand D. and Wang, Jiun-jie and Tsai, Chih-Chien and Druzgal, Jason and Newman, Benjamin T. and Melzer, Tracy R. and Pitcher, Toni L. and Dalrymple-Alford, John C. and Anderson, Tim J. and Garraux, Gaetan and Rango, Mario and Schwingenschuh, Petra and Suette, Melanie and Parkes, Laura M. and Al-Bachari, Sarah and Klein, Johannes and Hu, Michele T. M. and McMillan, Corey T. and Piras, Fabrizio and Vecchio, Daniela and Pellicano, Clelia and Zhang, Chengcheng and Poston, Kathleen L. and Ghasemi, Elnaz and Cendes, Fernando and Yasuda, Clarissa L. and Tosun, Duygu and Mosley, Philip and Thompson, Paul M. and Jahanshad, Neda and Owens-Walton, Conor and d'Angremont, Emile and van Heese, Eva M. and Laansma, Max A. and Altmann, Andre and Weil, Rimona S. and Oxtoby, Neil P.},
    title = {Neuroimaging-based data-driven subtypes of spatiotemporal atrophy due to {Parkinson's} disease},
    journal = {Brain Communications},
    volume = {7},
    number = {2},
    pages = {fcaf146},
    DOI = {10.1093/braincomms/fcaf146},
    year = {2025},
    type = {Journal Article}
 }

@manual{iit_atlas,
  title  = {IIT Human Brain Atlas v5.0},
  author = {Arfanakis, Konstantinos},
  year   = {2018},
  url    = {www.nitrc.org/projects/iit}
}

@article{10_stebbins2013identify,
  title={How to identify tremor dominant and postural instability/gait difficulty groups with the movement disorder society unified {Parkinson's} disease rating scale: comparison with the unified {Parkinson's} disease rating scale},
  author={Stebbins, Glenn T and Goetz, Christopher G and Burn, David J and Jankovic, Joseph and Khoo, Tien K and Tilley, Barbara C},
  journal={Movement Disorders},
  volume={28},
  number={5},
  pages={668--670},
  year={2013},
  publisher={Wiley Online Library}
}

@article{11_pySuStaIn,
   author = {Aksman, Leon M. and Wijeratne, Peter A. and Oxtoby, Neil P. and Eshaghi, Arman and Shand, Cameron and Altmann, Andre and Alexander, Daniel C. and Young, Alexandra L.},
   title = {pySuStaIn: A Python implementation of the Subtype and Stage Inference algorithm},
   journal = {SoftwareX},
   volume = {16},
   pages = {100811},
   ISSN = {2352-7110},
   DOI = {10.1016/j.softx.2021.100811},
   url = {https://www.sciencedirect.com/science/article/pii/S2352711021001096},
   year = {2021},
   type = {Journal Article}
}

@article{12_ENIGMA_PD,
   author = {Laansma, M. A. and Bright, J. K. and Al-Bachari, S. and Anderson, T. J. and Ard, T. and Assogna, F. and Baquero, K. A. and Berendse, H. W. and Blair, J. and Cendes, F. and Dalrymple-Alford, J. C. and de Bie, R. M. A. and Debove, I. and Dirkx, M. F. and Druzgal, J. and Emsley, H. C. A. and Garraux, G. and Guimaraes, R. P. and Gutman, B. A. and Helmich, R. C. and Klein, J. C. and Mackay, C. E. and McMillan, C. T. and Melzer, T. R. and Parkes, L. M. and Piras, F. and Pitcher, T. L. and Poston, K. L. and Rango, M. and Ribeiro, L. F. and Rocha, C. S. and Rummel, C. and Santos, L. S. R. and Schmidt, R. and Schwingenschuh, P. and Spalletta, G. and Squarcina, L. and van den Heuvel, O. A. and Vriend, C. and Wang, J. J. and Weintraub, D. and Wiest, R. and Yasuda, C. L. and Jahanshad, N. and Thompson, P. M. and van der Werf, Y. D. and Study, E. NIGMA-Parkinson's},
   title = {International Multicenter Analysis of Brain Structure Across Clinical Stages of {Parkinson's Disease}},
   journal = {Mov Disord},
   volume = {36},
   number = {11},
   pages = {2583-2594},
   ISSN = {1531-8257 (Electronic)
0885-3185 (Linking)},
   DOI = {10.1002/mds.28706},
   url = {https://www.ncbi.nlm.nih.gov/pubmed/34288137},
   year = {2021},
   type = {Journal Article}
}

@article{13_Young_DPM_review,
   author = {Young, A. L. and Oxtoby, N. P. and Garbarino, S. and Fox, N. C. and Barkhof, F. and Schott, J. M. and Alexander, D. C.},
   title = {Data-driven modelling of neurodegenerative disease progression: thinking outside the black box},
   journal = {Nat Rev Neurosci},
   volume = {25},
   number = {2},
   pages = {111-130},
   ISSN = {1471-003x},
   DOI = {10.1038/s41583-023-00779-6},
   year = {2024},
   type = {Journal Article}
}

@article{14_Moravveji_DPM_review,
   author = {Moravveji, S. and Doyon, N. and Mashreghi, J. and Duchesne, S.},
   title = {A scoping review of mathematical models covering Alzheimer's disease progression},
   journal = {Front Neuroinform},
   volume = {18},
   pages = {1281656},
   ISSN = {1662-5196 (Print)
1662-5196},
   DOI = {10.3389/fninf.2024.1281656},
   year = {2024},
   type = {Journal Article}
}

@article{15_TauFlowNet,
title = {TauFlowNet: Revealing latent propagation mechanism of tau aggregates using deep neural transport equations},
journal = {Medical Image Analysis},
volume = {95},
pages = {103210},
year = {2024},
issn = {1361-8415},
doi = {https://doi.org/10.1016/j.media.2024.103210},
url = {https://www.sciencedirect.com/science/article/pii/S136184152400135X},
author = {Tingting Dan and Mustafa Dere and Won Hwa Kim and Minjeong Kim and Guorong Wu}
}

@misc{16_COMIND,
      title={Scalable Modeling of Nonlinear Network Dynamics in Neurodegenerative Disease}, 
      author={Daniel Semchin and Emile d'Angremont and Marco Lorenzi and Boris Gutman},
      year={2025},
      eprint={2508.10343},
      archivePrefix={arXiv},
      primaryClass={q-bio.QM},
      url={https://arxiv.org/abs/2508.10343}, 
}
